\documentclass[aps,amsmath,amssymb,prb,twocolumn,showpacs,floatfix,tightenlines,pdftex]{revtex4}
\usepackage{bm}

\usepackage{graphicx}
\graphicspath{{./img/}{./}}

\usepackage{subfigure}

\usepackage{hyperref}

\begin{document}

\title{The emergence of gauge invariance: the stay-at-home gauge versus  local--global duality.}

\author{J. Zaanen}
\email{jan@lorentz.leidenuniv.nl}
\affiliation{Instituut-Lorentz for Theoretical Physics, Universiteit Leiden, P. O. Box 9506, 2300 RA Leiden, The Netherlands}

\author{A.J. Beekman}
\affiliation{Instituut-Lorentz for Theoretical Physics, Universiteit Leiden, P. O. Box 9506, 2300 RA Leiden, The Netherlands}

\begin{abstract}
In condensed matter physics gauge symmetries other than the $U(1)$ of electromagnetism are of an emergent nature. Two emergence 
mechanisms for gauge symmetry are well established: the way these arise in Kramers--Wannier type local--global dualities, and as a way to
encode local constraints encountered in (doped) Mott insulators. We demonstrate that these gauge structures are closely related, and appear as counterparts in either the canonical or  field-theoretical language. The restoration of symmetry in a disorder phase transition is due to having the original local variables subjected to a coherent superposition of all possible topological defect configurations, with the effect that correlation functions are no longer well-defined. This is completely equivalent to assigning gauge freedom to those variables. Two cases are considered explicitly: the well-known vortex duality in bosonic Mott insulators serves to illustrate the principle. The acquired wisdoms are then applied to the less familiar context of dualities in quantum elasticity,
where we elucidate the relation between the quantum nematic  and linearized gravity. We reflect on some deeper implications for  the emergence of gauge symmetry in general.
\end{abstract}

\date{August 13, 2011}

 \maketitle

\section{Introduction.}

Breaking symmetry is easy but making symmetry is hard: this wisdom applies to global symmetry 
but not to local symmetry. The study of systems controlled by emergent gauge symmetry has become
a mainstream in modern condensed matter physics. Although one discerns only the gauge invariance of electromagnetism in the ultraviolet of condensed matter physics, it is now very well understood that 
in a variety of circumstances  gauge symmetries that do not exist on the microscopic scale control the 
highly collective physics on the macroscopic scale. An intriguing but unresolved issue is whether
the gauge structures involved in the Standard Model of high energy physics and perhaps even general 
relativity could be of such an emergent kind. 

The mechanisms for gauge symmetry emergence fall into two broad categories: (1) the ``stay-at-home" gauge 
invariance associated with (doped) Mott insulators and the gauge fields associated with the slave-particle theories
encountered in this context, describing the fractionalization
of the quantum numbers of the microscopic degrees of freedom, and (2) the global-to-local symmetry correspondence encountered
in the strong--weak (Kramers--Wannier, $S$-) dualities as of relevance to the quantum field theories describing the
collective quantum physics of condensed matter systems. In the common perception these appear as quite  
different. The purpose of this note is to clarify that at least in the context of bosonic physics they are actually
closely related.  In fact, these highlight complementary aspects of the vacuum structure, and it is just pending
whether one views the vacuum either using the canonical/Hamiltonian language (stay-at-home) or field-theoretical/Lagrangian
(local--global duality) language. The case is very simple and we will illustrate it in section \ref{sec:Vortex duality versus Bose Mott insulators} with the most primitive of all 
many-particle systems governed by continuous symmetry: the Bose-Hubbard model at zero chemical potential, 
or alternatively the Abelian-Higgs duality associated with complex-scalar field theory. Although we do not claim 
any new result in this particular context, the freedom to switch back and forth between the Lagrangian and 
Hamiltonian viewpoints yields some entertaining vistas on this well-understood theory. To make the case that it
can yield new insight, we apply it in section \ref{sec:Quantum nematic crystals and the emergence of linearized gravity} to the less familiar context of dualities in quantum elasticity. This deals with
the description of quantum liquid crystals in terms of dual condensates formed from the translational topological 
defects (dislocations) associated with the fully ordered crystal. Using the Lagrangian language it was argued 
that such quantum nematics are equivalent to (linearized) Einstein gravity \cite{KleinertZaanen04}. Here we will demonstrate that 
this is indeed controlled by the local symmetry associated with linearized gravity: translations are gauged, turning into infinitesimal 
Einstein transformations. 

This could have been a very short communication but we wish to address a readership with diverse backgrounds. We therefore 
first review at length the basics of Mottness (section \ref{sec:stay-at-home gauge}) and vortex duality (section \ref{sec:vortex duality}) which should be quite familiar for the condensed matter
physicists, presenting in section \ref{sec:The vortex condensate as the generator of the hidden stay-at-home gauge} our argument revealing how the stay-at-home gauge is encoded in the vortex duality. Section \ref{sec:Quantum nematic crystals and the emergence of linearized gravity}
is devoted to duality in quantum elasticity and the relation to gravity. We again take the time to review the basics since this subject is unfamiliar even for the
condensed matter mainstream. In section \ref{sec:The vortex condensate as the generator of the hidden stay-at-home gauge} we review the basics of these duality structures while in section \ref{sec:The field theory of quantum elasticity and geometry: the Kleinert rules.} we turn to the way this is
related to gravitational physics, employing the insights derived from vortex duality to show why the effective space realized in the quantum nematic
behaves like the spacetime of general relativity when it is nearly flat.

\section{Vortex duality versus Bose-Mott insulators.}\label{sec:Vortex duality versus Bose Mott insulators}

In this section we will first review two standard views on the physics of the  Bose-Hubbard model \cite{FisherEtAl89}, with the
intention to confuse the reader. We first highlight the emergence of the compact $U(1)$ stay-at-home
gauge which emerges in the Bose-Mott insulator in a trivial way when one sticks to the canonical, second
quantized language focusing on the large-$U$ limit. In the second subsection we turn to the field-theoretical
Abelian-Higgs duality, reviewing the standard argument that in 2+1D the quantum disordered partner of the
superfluid is actually dual to a superconducting condensate formed from the vortices of the superfluid, which interact
via effective $U(1)$ gauge fields \cite{FisherLee89,Kleinert89a,HerbutTesanovic96,CvetkovicZaanen06a,NguyenSudbo99,HoveSudbo00}.  These should be equivalent descriptions of the Bose-Mott insulator, but at
face value this is far from obvious. In the last subsection we present a very simple resolution of this conundrum,
having some interesting ramifications for the way one should think in general about ``Mott insulators" in field theory.    

\subsection{The stay-at-home gauge of the Bose-Mott insulator.}\label{sec:stay-at-home gauge}

A mainstream of the gauge theories in condensed matter physics dates back to the late 1980s when
the community was struggling with the fundamentals of the problem of high-$T_\mathrm{c}$ superconductivity. It 
was recognized early on that this has to do with doping the parent Mott insulators and this revived 
the interest in the physics of the Mott insulating state itself \cite{LeeNagaosaWen06,Phillips10,Zaanen11,MrossSenthil11}. The point of departure is the Hubbard 
model for electrons,

\begin{equation}
H_\mathrm{FH} = -t \sum_{<ij>\sigma} (\hat{c}^{\dagger}_{i \sigma} \hat{c}_{j \sigma} + \hat{c}^{\dagger}_{j \sigma} \hat{c}_{i \sigma} )  + U \sum_{i} \hat{n}_{i\uparrow} \hat{n}_{i\downarrow} \; ,
\label{Hubbardfermions}
\end{equation}

describing fermions $\hat{c}^\dagger_{i \sigma}$ on site $i$ with spin $\sigma$, hopping on a lattice with rate $t$, subjected to a strong local Coulomb interaction $U$. Here $\hat{n}_{i \sigma} = \hat{c}^\dagger_{i \sigma} \hat{c}_{i \sigma}$ is the fermion number operator. When $U \gg t$ 
and when there is on average one fermion per site, it is easy to recognize the Mott insulator: a fermion will localize at
every site, while  it costs an energy $\sim U$  to move a charge which will therefore not happen at macroscopic scales.
This simplicity is deceptive: viewed from a general perspective the effect of this ``projective renormalization" is profound. 
Due to the dynamics as dominated by  the strong local repulsions the particle number becomes locally conserved. In the large-$U$ limit,

\begin{equation}
\sum_{\sigma} \hat{n}_{i\sigma} | \Psi \rangle = | \Psi \rangle \;,
\label{numberquant}
\end{equation}

where $|\Psi \rangle$ is the ground state wave function.   Although the on-site fermion number is not a conserved quantity of 
the microscopic Hamiltonian Eq. (\ref{Hubbardfermions}) it becomes locally conserved at low energy.  
This truly emergent  local constraint/conservation law can be imposed by a compact $U(1)$ gauge field $\alpha_{i\sigma}$,

\begin{eqnarray}
\hat{c}^{\dagger}_{i\sigma} & \rightarrow &   \hat{c}^{\dagger}_{i\sigma} e^{i\alpha_{i\sigma}}\;, \nonumber \\
\hat{c}_{i\sigma} & \rightarrow &  e^{- i\alpha_{i\sigma}} \hat{c}_{i\sigma}\;, \nonumber \\
\hat{n}_i  & = & \sum_{\sigma} \hat{c}^{\dagger}_{i\sigma} \hat{c}_{i\sigma} \rightarrow \hat{n}_i\;.
\label{numberphasefer}
\end{eqnarray}

In this effortless way one discovers that a gauge symmetry emerges that controls the physics at long distances,
while it is non existent at the microscopic scale. This is the point of departure of a mainstream  school of thought
in condensed matter physics.   There is still a dynamical spin system at work at low energies. Using various ``slave-constructions" it was
subsequently argued that quantum spin liquids characterized by fractionalized excitations can be realized
when the resulting compact $U(1)$ gauge theory would end up in a  deconfining regime. 

A much simpler problem is the Bose-Hubbard model describing spinless  bosons created by $\hat{b}^\dagger_i$ hopping on a lattice
with a rate $t$ subjected to an on-site repulsion $U$,   

\begin{equation}
H_\mathrm{BH} = -t \sum_{<ij>} \hat{b}^{\dagger}_i  \hat{b}_j  + U \sum_{i} \hat{n}_i^2 \;.
\label{Hubbardbosons}
\end{equation}

Again $\hat{n}_i = \hat{b}^\dagger_i \hat{b}_i$ is the boson number operator. We assume in the remainder that the system is at ``zero chemical potential", meaning that on average 
there is an integer number of bosons $n_0$ per site. Recently this problem attracted attention
in the context of cold bosonic atoms  on a optical lattice \cite{GreinerEtAl02}, while it describes equally well Josephson junction
networks\cite{FazioVanderZant01}. In the large-$U$ limit one obtains the same simple picture for the Bose-Mott insulator
as for its fermionic sibling, except that the spinless Bose  variety is seen as completely featureless 
since it does not seem to break a manifest symmetry while low energy degrees of freedom are absent. However, it does share the trait 
with the fermionic Mott insulator that  when $U/t  \gg 1$ a low energy 
compact $U(1)$ gauge invariance is generated since the number of bosons per site is sharply quantized,

\begin{eqnarray}
\hat{b}^{\dagger}_i & \rightarrow &   \hat{b}^{\dagger}_i e^{i\alpha_i} \nonumber \;,\\
\hat{b}_i & \rightarrow &  e^{- i\alpha_i} \hat{b}_i\;, \nonumber \\
\hat{n}_i  & = & \hat{b}^{\dagger}_i  \hat{b}_i \rightarrow \hat{n}_i\;.
\label{numberphase}
\end{eqnarray}
  
In 2+1 and higher dimensions it is well established that the model Eq. (\ref{Hubbardbosons}) is characterized by a
quantum phase transition as function of the coupling  $g = U/n_0 t$ at a critical coupling $g_\mathrm{c}$ . The Mott insulator 
is established  for $g > g_\mathrm{c}$ while for small couplings $g < g_\mathrm{c}$ the zero temperature state is a superfluid. In this 
regime it is convenient to rewrite the Hamiltonian in phase representation, $\hat{b}^{\dagger}_i = | b | e^{i\hat{\phi}_i}$, 
to obtain the phase dynamics model ($J = n_0 t$), 

\begin{equation}
H_{\phi} = J \sum_{<ij>}  \cos (\hat{\phi}_i - \hat{\phi}_j) +U \sum_i  \hat{n}_i^2\;,
\label{phasedynH}
\end{equation}

subjected to the quantization condition $[ \hat{\phi}_i , \hat{n}_i ] = i \delta_{ij}$.  For $g < g_c$ the global $U(1)$ symmetry
associated with the phase $\phi$ will break spontaneously and this results in the superfluid. In terms of the original boson the
symmetry breaking implies that it develops a vacuum expectation value (VEV),

\begin{equation}
\langle \hat{b}^{\dagger}_i \rangle \rightarrow  \sqrt{n_0} e^{i\phi_0}\;,
\label{VEVboson}
\end{equation}

where $\phi_0$ the phase shared by the condensate, and $n_0$ the average density.  Comparing with Eq. (\ref{numberphase}), 
it is obvious that the global $U(1)$ symmetry that breaks spontaneously in the phase condensate acquires an emergent compact $U(1)$
gauge status in the ``number condensate" realized in the Mott insulator. This Bose-Hubbard model reveals perhaps the most primitive mechanism
for the emergence of a local symmetry from microscopic global symmetry. 

\subsection{Phase dynamics duality: the superconductor as the dual of the superfluid.}\label{sec:vortex duality}

Let us now turn to the field-theoretical formulation of the same problem. The quantum partition sum is given in terms of the Euclidean path
integral,

\begin{equation}
Z = \int {\cal{D}} \phi\ e^{ \frac{1}{g} \int^{\beta}_0 d\tau d^d x\ {\cal{L}} }
\label{partsumphase}
\end{equation}

The Lagrangian $\cal{L}$ is derived from Eq. (\ref{phasedynH}) departing from the superfluid phase
by realizing that in the Legendre transformation $\hat{n}_i \rightarrow i \partial_{\tau} \hat{\phi}_i$, 
followed by a naive coarse graining. After scaling out the phase velocity $c_{\mathrm{ph}} \sim \sqrt{ UJ}$
and defining the coupling constant as $g = U/J$, one finds,

\begin{equation}
{\cal{L}} = \frac{1}{2g}\left( \partial_{\mu} \phi (\vec{x}) \right)^2\;,
\label{phaselag}
\end{equation}

in a  relativistic short hand notation ($\vec{x}, \mu = \tau, x, y, ...$), while the field $\phi$ is compact 
with periodicity $2\pi$.  This theory just describes the Goldstone boson/phase mode of the ordered superfluid. 
Recently we showed that this theory is subjected to a global--local duality in all dimensions equal to and larger than 2+1D 
  \cite{BeekmanSadriZaanen11}. Here we will just consider  the familiar 
 2+1D ``Abelian-Higgs" or ``vortex" duality case since it can be regarded in the present context as fully 
 representative for the higher dimensional cases as well. One first transforms
Eqs. (\ref{partsumphase},{\ref{phaselag}) to vortex  coordinates. After a Hubbard--Stratonovich transformation,
${\cal{L}} = \frac{g}{2} J^2_{\mu} + i J_\mu \partial_{\mu} \phi$. The auxiliary fields $J_{\mu}$ just represent the supercurrents
dual to the phase field. This is easily seen by factorizing the phase field in smooth and multivalued configurations,
$\phi = \phi_{\mathrm{sm}} + \phi_{\mathrm{MV}}$. The smooth $\phi_{\mathrm{sm}}$ enters as a Lagrange multiplier which can be integrated 
out yielding the conservation law $\partial_{\mu} J_{\mu} = 0$, which is just the supercurrent continuity equation.
In 2+1D this can be imposed by parametrizing the current in terms of a {\em non-compact} $U(1)$ gauge field
$A_{\mu}$ as $J_{\mu} = \epsilon_{\mu \nu \lambda} \partial_{\nu} A_{\lambda}$. Inserting this in the action,
one finds after some straightforward reshufflings,

\begin{eqnarray}
\cal{L} & =  & \frac{g}{4}F_{\mu \nu} F_{\mu \nu} + i A_{\mu} J^\mathrm{V}_{\mu}\;, \nonumber \\
J^\mathrm{V}_{\mu} & =  & \epsilon_{\mu \nu \lambda} \partial_{\nu} \partial_{\lambda} \phi_\mathrm{MV} \;,
\label{vortexlag}
\end{eqnarray}

where $F_{\mu \nu} = \partial_{\mu} A_{\nu} - \partial_{\nu} A_{\mu}$, the field strength of $A_{\mu}$ 
which is now sourced by vortex currents; one recognizes that the non-integrability in the phase field  
$J^\mathrm{V}_{\mu}$ just corresponds with the vorticity.

We encounter here yet another emergent gauge theory, but this one is completely unrelated to the gauging 
of the phase field in the canonical formulation. We have just written down an action describing how the 
topological excitations of the superfluid  (vortices) mutually interact. Since the phase field deformation
due to the vortex is long ranged, the vortex--vortex interaction is long ranged as well. The above derivation shows
that these long range vortex interactions are actually indistinguishable from the electromagnetic interactions between 
electrically charged particles in 2+1D. It is in a way just a convenience that these can be parametrized in terms of 
gauge potentials, and these {\em non-compact} $U(1)$ fields have clearly no relation to the emergent {\em compact}
$U(1)$ fields imposing number quantization in the Mott insulating phase.

The next step in vortex duality is to consider what happens when the coupling constant $g$ increases. The vortices 
are the unique sources of quantum fluctuations and for small couplings they occur as small vortex--anti-vortex loops
in Euclidean spacetime. For growing coupling constant these loops will become larger until at $g_\mathrm{c}$ a ``loop blowout"
occurs where they become infinitely large. This is the quantum phase transition, where at larger couplings a ``tangle of free
(anti-)vortex world lines'' is found in spacetime.  This Bose condensate of vortices just corresponds with the quantum disordered
phase, and because of the gauge interactions between the vortices this is just the same entity as a condensate of relativistic bosons interacting via electromagnetic gauge fields. This condensate is governed by the Ginzburg--Landau--Wilson action,

\begin{widetext}
\begin{equation}
S_{V} = \int d\tau \int dx^2\ \frac{1}{2}\left[  |(\partial_{\mu} + i A_{\mu}) \Psi_\mathrm{V}|^2 + m_\mathrm{V}^2 |\Psi_\mathrm{V}|^2 + \frac{1}{2}w_\mathrm{V} |\Psi_\mathrm{V}|^4
+ \frac{g}{2} F_{\mu \nu} F_{\mu \nu} \right]\;,
\label{vortexGLW}
\end{equation}
\end{widetext}

where $\Psi_\mathrm{V}$ is the vortex condensate field, while the gauge fields $A_{\mu}$ descend from Eq. (\ref{vortexlag}).
This is in turn just   a relativistic {\em superconductor} or 
``Higgs phase". In the condensate the vortex-matter field can be integrated out and the effective action for the 
massive photons can be written in a gauge invariant fashion in terms of the original supercurrents $J_{\mu}$ as,

\begin{equation}
S_{\mathrm{Mott}} = \int d\tau \int dx^2\ \left[  m^2_{\mathrm{Mott}} J_{\mu} \frac{1}{\partial^2} J_{\mu} + J_{\mu} J_{\mu} \right]
\label{vmassphot}
\end{equation}

Describing a doublet of propagating photons with a mass $m_{\mathrm{Mott}}$.  
The conclusion is that in 2+1D a quantum disordered superfluid (in terms of phase) can be 
equally well be interpreted as a relativistic  superconductor formed from the vortices.  

\subsection{The vortex condensate as the generator of the hidden stay-at-home gauge.}\label{sec:The vortex condensate as the generator of the hidden stay-at-home gauge}

Up to this point we have just collected and reviewed some well-known results on phase dynamics. However, 
at first sight it might appear as if the matters discussed in the two previous subsections are completely
unrelated. Departing from the Bose-Hubbard model the considerations of the previous subsection leave no
doubt that in one or the other way the dual vortex superconductor can be adiabatically continued all the
way to the strongly coupled Bose-Mott insulator of the first subsection. The standard way to argue this is
by referral to the excitation spectrum. The Bose-Mott insulator is characterized by a mass gap $\sim U$
(at strong coupling), and a doublet of ``holon" (vacancy) and ``doublon" (doubly-occupied site) propagating
excitations being degenerate at zero chemical potential. The vortex superconductor is a
relativistic $U(1)/U(1)$  Higgs condensate in 2+1 dimensions characterized by a Higgs mass (a gap) above 
which one finds a doublet of spin-1 (left and right helical) ``vector bosons". In this regard there is a precise
match. However, in the canonical formalism one also discovers the emergent $U(1)$ invariance associated with the 
sharp quantization of local number density in the Mott insulator. What has happened with this important
symmetry principle in the vortex superconductor? 

The answer is:  the emergent compact $U(1)$ gauge symmetry of the Mott insulator 
is actually a generic part of the physics of the relativistic superconductor.  

The argument is exceedingly simple. The stay-at-home gauge does not show
up explicitly in the  Higgsed action describing the dual vortex condensate, for the simple reason that all the 
quantities in this action are associated with the vortices which are in turn in a perfect non-local relation with 
the original phase variables. However, we know precisely what this dual superconductor is in  terms of
those phase variables. We can resort to a first quantized, world line description of the vortex superconductor, putting
back ``by hand" the phase variables. This constitutes a tangle of world lines of vortices, warping the original phases,
and eventually we can even map that back to a first quantized wave function written as a coherent superposition
of configurations of the phase field. To accomplish this in full one needs big computers \cite{NguyenSudbo99,HoveSudbo00}, but 
for the purposes of scale and symmetry analysis the outcomes are obvious.

The penetration depth $\lambda_{\mathrm{V}}$ of the dual vortex superconductor is just 
coincident with the typical distance between vortices. At distances much shorter than $\lambda_{\mathrm{V}}$ the vortices do not
scramble the relations between the phases at spatially separated points and at these scales the system behaves as
the ordered superfluid,

\begin{equation}
\langle b^{\dagger} (r) b (0) \rangle \rightarrow {\mathrm{constant}}\;, \quad r \ll \lambda_\mathrm{V}\;,
\label{noMott}
\end{equation}

\begin{figure*}
\huge
 $\frac{1}{\sqrt{2}}\Big( |$  \raisebox{-8mm}{\includegraphics[height=2cm]{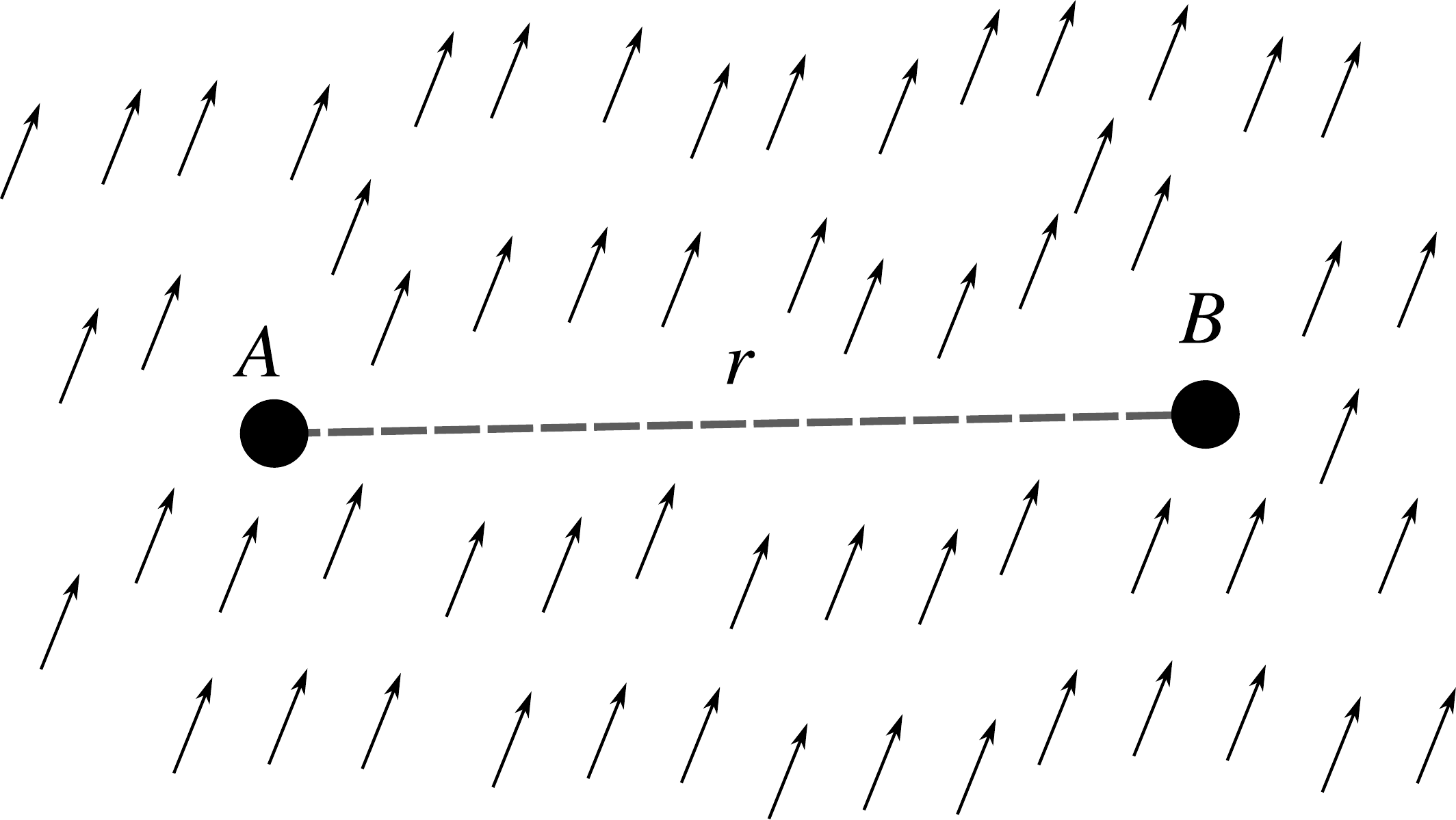}} $\rangle + |$  \raisebox{-8mm}{\includegraphics[height=2cm]{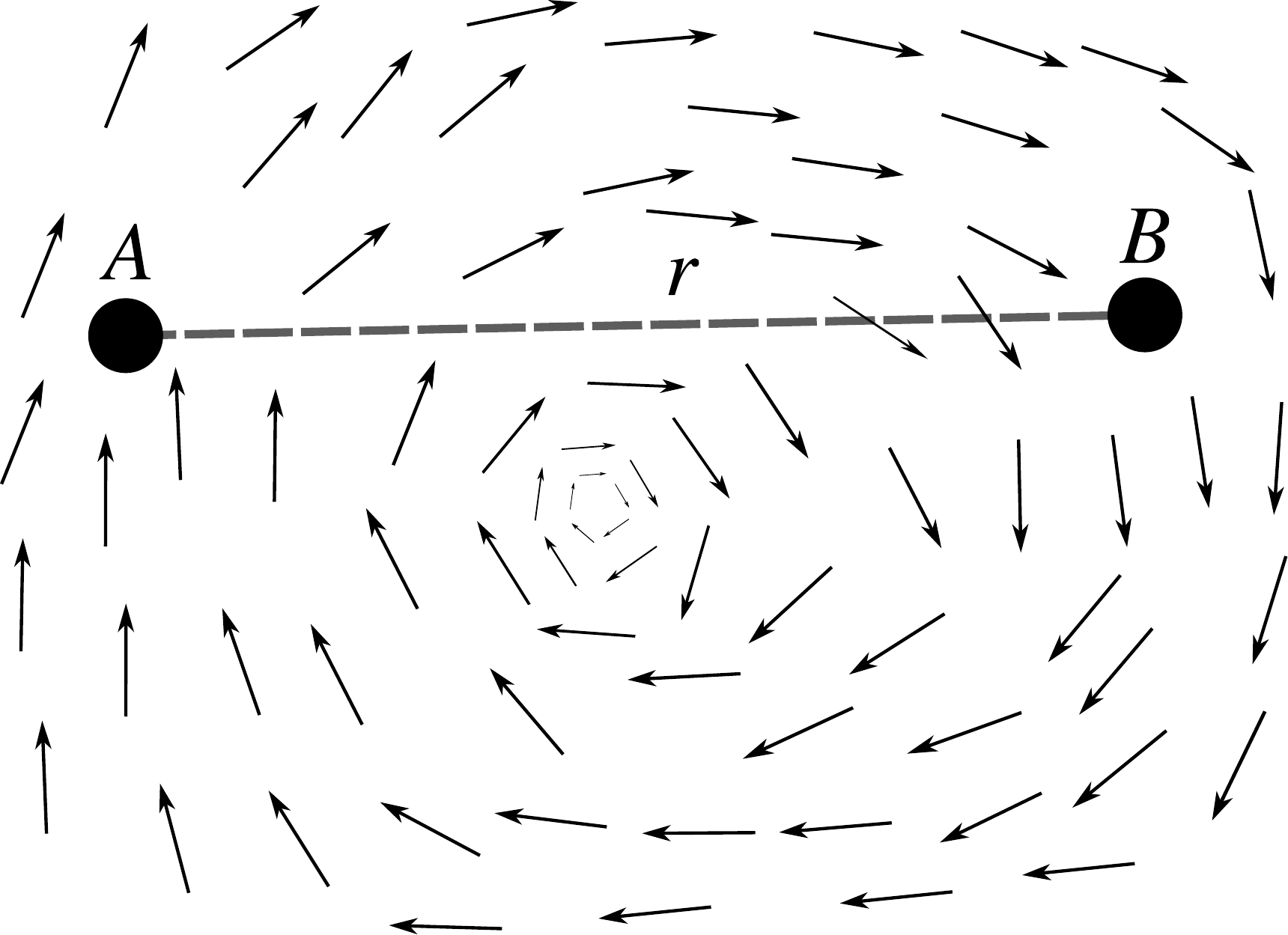}} $\rangle \Big)$
\normalsize
\caption{In the vortex condensate the correlation of the phase between a point $A$ and another point $B$ a distance $r$ apart is in a superposition of have zero, one or any number of vortices in between. As such the phase at $B$ with respect to that at $A$ is completely undefined: it has acquired a complete gauge invariance in the sense that any addition to phase is an equally valid answer\label{fig:phase superposition}}
\end{figure*}

However, at distances of order $\lambda_{\mathrm{V}}$ and larger, the vacuum turns into a coherent quantum  superposition of 
``Schr\"odinger cat states" where there is either none, or one,  or whatever number of vortices in between the 
points $0$ and $r$ where we ask the question of  how to correlate the phases of  bosons, see Fig. \ref{fig:phase superposition}. 
We have arrived at exposing the simple principle which is the central result of this paper: {\em since the vortex configurations are in coherent superposition, the phases acquire a full compact $U(1)$ 
gauge invariance.} The computation is already done using the canonical formalism, and all that remains is to understand
the physical concept: focus on the direction of the phase at the origin and look at the phase arrow at some distance point
$r$. Consider a particular configuration of the vortices, and in this realization the distant phase will point in some definite direction 
which will be different from the phase at the origin as determined by  the particular vortex configuration.  However, since all different vortex configurations 
are in coherent superposition and therefore ``equally true at the same time",  all orientations of the phase at point $r$ are
also ``equally true at the same time" and this is just the precise way to formulate that a compact $U(1)$ gauge symmetry
associated with $\phi$ has emerged at  distances  $\lambda_\mathrm{V}$. 

The implication is  that via Eq. (\ref{numberphase}) the emerging stay-at-home gauge invariance implies that 
in the Higgs condensate the number density associated with the bosons condensing in the dual superfluid  
becomes locally conserved on the scale $\lambda_V$. The  Mottness  therefore sets in only at scales larger than this $\lambda_V$. 
Notice that this mechanism does in fact not need a lattice: it is just a generic property of the field theory itself, which is independent 
of regularization. In fact, the seemingly all important role of the lattice in the standard reasoning in condensed matter
when dealing with these issues is a bit of tunnel vision. It focuses on the strong-coupling limit where for large 
$U$, $\lambda_\mathrm{V} \rightarrow a$, the lattice constant. However, upon decreasing the coupling the stay-at-home gauge
emerges at an increasingly longer length scale $\lambda_\mathrm{V}$, to eventually diverge at the quantum phase transition. 
Close to the quantum critical point the theory has essentially forgotten about the presence of the lattice, just remembering
that it wants to conserve number locally which is the general criterium to call something an insulator. In fact, Mottness can exist
without a lattice altogether. A relativistic superconductor living in a perfect 2+1D continuum is physically reasonable. Since duality
works in both directions, this can be in turn viewed as a quantum disordered superfluid, where the number density associated 
with the bosons comprising the superfluid becomes locally conserved. 

By inspecting closely this simple vortex duality we have discovered a principle which might be formulated in full generality as: {\em the coherent
superposition of the disorder operators associated with the condensation of the disorder fields has the automatic consequence that 
the order fields acquire a gauge invariance associated with the local quantization of the operators conjugate to the operators 
condensing  in the order field theory.}  We suspect that this principle might be of use also in the context of dualities involving more complex
field theories. To substantiate this claim, let us now inspect  a more involved duality which is encountered in quantum elasticity, where 
the principle reveals the precise reasons for why quantum liquid crystals have dealings with general relativity. 

 \section{Quantum nematic crystals and the emergence of linearized gravity.}\label{sec:Quantum nematic crystals and the emergence of linearized gravity}

Einstein himself already forwarded the metaphor that the spacetime of general relativity is like an elastic medium. 
Is there a more literal truth behind this metaphor? In recent years the mathematical physicist Hagen Kleinert has been forwarding the view that quite deep analogies exist between plastic media (solids with topological defects) and 
Einsteinian spacetime \cite{Kleinert89b,Kleinert08}.  There appears room for the possibility that at the Planck scale an exotic ``solid" (the ``world 
crystal") is present, turning after coarse graining into the spacetime that we experience. 

It turns out that this subject matter has some bearing on a much more practical question: what is the general nature 
of the quantum hydrodynamics and rigidity of quantum liquid crystals? Quantum liquid crystals\cite{FradkinEmeryKivelson98} are just the zero 
temperature versions of the classical liquid crystals found in computer displays. These are substances characterized by a  partial breaking of spatial symmetries, while the zero temperature versions are at the same time quantum liquids.  Very recently indications have been found  for variety of such quantum liquid crystals in experiment\cite{FradkinKivelson99,HinkovEtAl08,DaouEtAl10,LawlerEtAl10,BorziEtAl07,ChuEtAl10}. In the present context we are especially interested in the ``quantum  smectics" and ``quantum nematics" found in high-$T_\mathrm{c}$ cuprates\cite{HinkovEtAl08,DaouEtAl10,LawlerEtAl10,MesarosEtAl11} 
which appear to be also superconductors at zero temperature.  Such matter should be at least in the long-wavelength limit governed by a bosonic field theory, and this ``theory of quantum elasticity"  \cite{ZaanenNussinovMukhin04,CvetkovicNussinovMukhinZaanen08,CvetkovicZaanen06b,Cvetkovic06} is characterized by dualities that are richer, but eventually closely 
related to the duality discussed in the previous section. Departing from the quantum crystal, the topological agents which are responsible for the restoration of symmetry are the dislocations and disclinations. The disclinations restore the rotational symmetry and the topological criterium for liquid crystalline order is that these continue to be massive excitations. The dislocations restore translational symmetry, and these are in crucial regards similar to the vortices
of the previous section. In direct analogy with the Mott insulator being a vortex superconductor, the superconducting 
smectics and nematics can be universally viewed as dual ``stress superconductors"  associated with Bose condensates of quantum dislocations. 

Using the geometrical correspondences of Kleinert \cite{Kleinert89b,Kleinert08}, arguments were forwarded  suggesting  
that the Lorentz-invariant version of the superfluid nematic in 2+1D is characterized by a low energy dynamics that is the same as at least linearized gravity\cite{KleinertZaanen04}. Very recently it was pointed out that this appears also to be the case in the 3+1D case\cite{ZhuJiang11,BeekmanSadriZaanen12}. A caveat 
is that Lorentz invariance is badly broken in the liquid crystals as realized in condensed matter physics. This changes the rules drastically and although 
the consequences are well understood in 2+1D\cite{CvetkovicNussinovMukhinZaanen08,CvetkovicZaanen06b,Cvetkovic06} it remains to be clarified what this means 
for the 3+1D condensed matter quantum liquid crystals. The unresolved issue is how the gravitons of the 3+1D relativistic case imprint on the collective modes of the non-relativistic systems.

Here we want to focus on perhaps the most fundamental question one can ask in this context: although general relativity is not a Yang--Mills theory, it is uniquely associated with the gauge symmetry of general covariance or diffeomorphism. Quite generally, attempts to identify ``analogue" or ``emergent" gravity in 
condensed matter systems have been haunted  by the problem that general covariance is quite unnatural in this context.  The gravity analogues in so far 
identified in condensed matter like systems usually get so far that a non-trivial geometrical parallel transport  of the matter is identified, 
that occurs in a ``fixed frame" or ``preferred metric"\cite{Volovik01,Volovik03,Volovik04,Visser98,Garay2000,Jannes09,Chapline03,MesarosSadriZaanen10}, 
or either this issue of the mechanism of emerging general covariance is just not addressed\cite{GuWen06,ZhangHu01,WuWangGeYang10}.  
As we will discuss, crystals are manifestly non-diffeomorphic.  However, the relativistic quantum nematics appear to be dynamically similar to
Einsteinian spacetime. For this to be true, it has to be that in one or the other way general covariance is emerging in such systems.  How does this
work?

In close parallel with the vortex duality ``toy model" of the previous section, we will  explicitly demonstrate in this section that 
indeed general covariance is dynamically generated as an emergent IR symmetry. However, there is a glass ceiling: the geometry is only 
partially gauged. Only the infinitesimal ``Einstein" translations fall prey to an emergent gauge invariance while the Lorentz transformations remain in a fixed 
frame. This prohibits the identification of black holes and so forth, but this symmetry structure turns out to be 
coincident with the gauge fix that is underlying linearized gravity. The conclusion is that relativistic quantum nematics form a medium that supports gravitons, but nothing else than gravitons. 

For this demonstration we have to rely on the detour for the identification of the local symmetry generation as introduced in the previous section. 
Different from the  Bose-Mott insulator, there is no formulation available for the quantum nematic in terms of a simple Hamiltonian where one can 
directly read off the equivalent of the stay-at-home gauge symmetry.  We have therefore to find the origin of the gauging of the Einstein translations 
in the physics of the dislocation Bose condensate, but this will turn out to be a remarkably simple and elegant affair. 
 
The remainder of this section is organized as the previous one. In section \ref{sec:Duality in quantum elasticity: the quantum nematic as a dislocation Bose condensate} we will first collect the various bits and pieces: a
sketch of the way that ``dislocation duality" associates the relativistic quantum nematic state with a crystal that is
destroyed by a Bose condensate of dislocations. In section \ref{sec:The field theory of quantum elasticity and geometry: the Kleinert rules.} we will subsequently review Kleinert's ``dictionary" relating 
quantum elasticity and Einsteinian geometry, while at the end of this subsection we present the mechanism of gauging Einstein translations by the dislocation condensate. For simplicity we will focus on the 2+1D case; the generalities we address here apply equally well to the richer 3+1D case. 

\subsection{Duality in quantum elasticity: the quantum nematic as a dislocation Bose condensate.}\label{sec:Duality in quantum elasticity: the quantum nematic as a dislocation Bose condensate}

Let us first introduce the field-theoretical side \cite{ZaanenNussinovMukhin04,CvetkovicNussinovMukhinZaanen08,CvetkovicZaanen06b,Cvetkovic06}. The theory of quantum elasticity is just the 19th century theory of 
elasticity but now embedded in the Euclidean spacetime of thermal quantum field theory. To keep matters
as simple as possible we limit ourselves to the Lorentz-invariant ``world crystal", just amounting to the 
statement that we are dealing with a 2+1D elastic medium being isotropic, both in space and time directions,

\begin{eqnarray}
Z & = & \int D w\ e^{- \frac{1}{2}S_\mathrm{el}}\;, \nonumber \\
S_\mathrm{el} & = & \int d\tau \int dx^2\ \left[ \mu w_{\mu \nu} w^{\mu \nu} + \frac{\lambda}{2} w^2_{\mu \mu} \right] \;,
\label{elfieldth}
\end{eqnarray}

where,

\begin{equation}
w_{\mu \nu} = \frac{1}{2} \left( \partial_{\mu} u_{\nu} + \partial_{\nu} u_{\mu} \right) \;,
\label{straindef}
\end{equation}

are the strain fields associated with the displacements $u_{\nu}$ of the ``world crystal atoms" relative to their 
equilibrium positions. As before $g$ is  the coupling constant, while $\mu$ and $\lambda$ are the shear modulus
and the Lam\'{e} constant  of the world crystal, respectively. At first view this looks like a straightforward tensorial 
generalization of the scalar field theory of the previous section. For the construction of the nematics one can indeed
think about the displacements as ``scalar fields with flavors" since this only involves the ``Abelian sector" of the theory associated 
with translations. One should keep in mind however that one is breaking Euclidean space down to a lattice subgroup
and this is associated with non-Abelian, infinite and semi-direct symmetry structure: the full theory beyond the
dislocation duality is a much more complicated affair. 

\begin{figure}
 \subfigure[Edge dislocation]{\includegraphics[width=4cm]{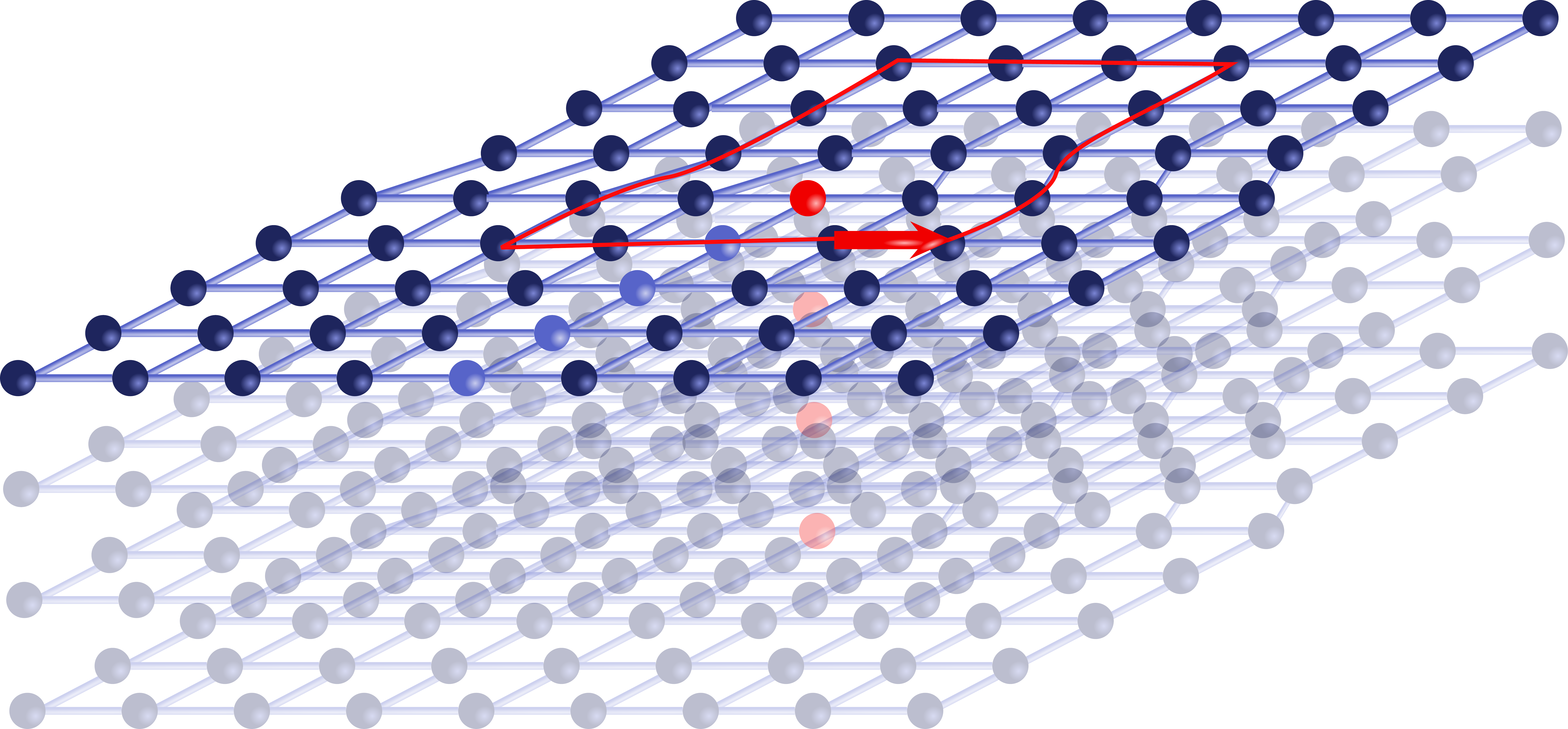}\label{fig:edge dislocation}}
\hfill
 \subfigure[Screw dislocation]{\includegraphics[width=4cm]{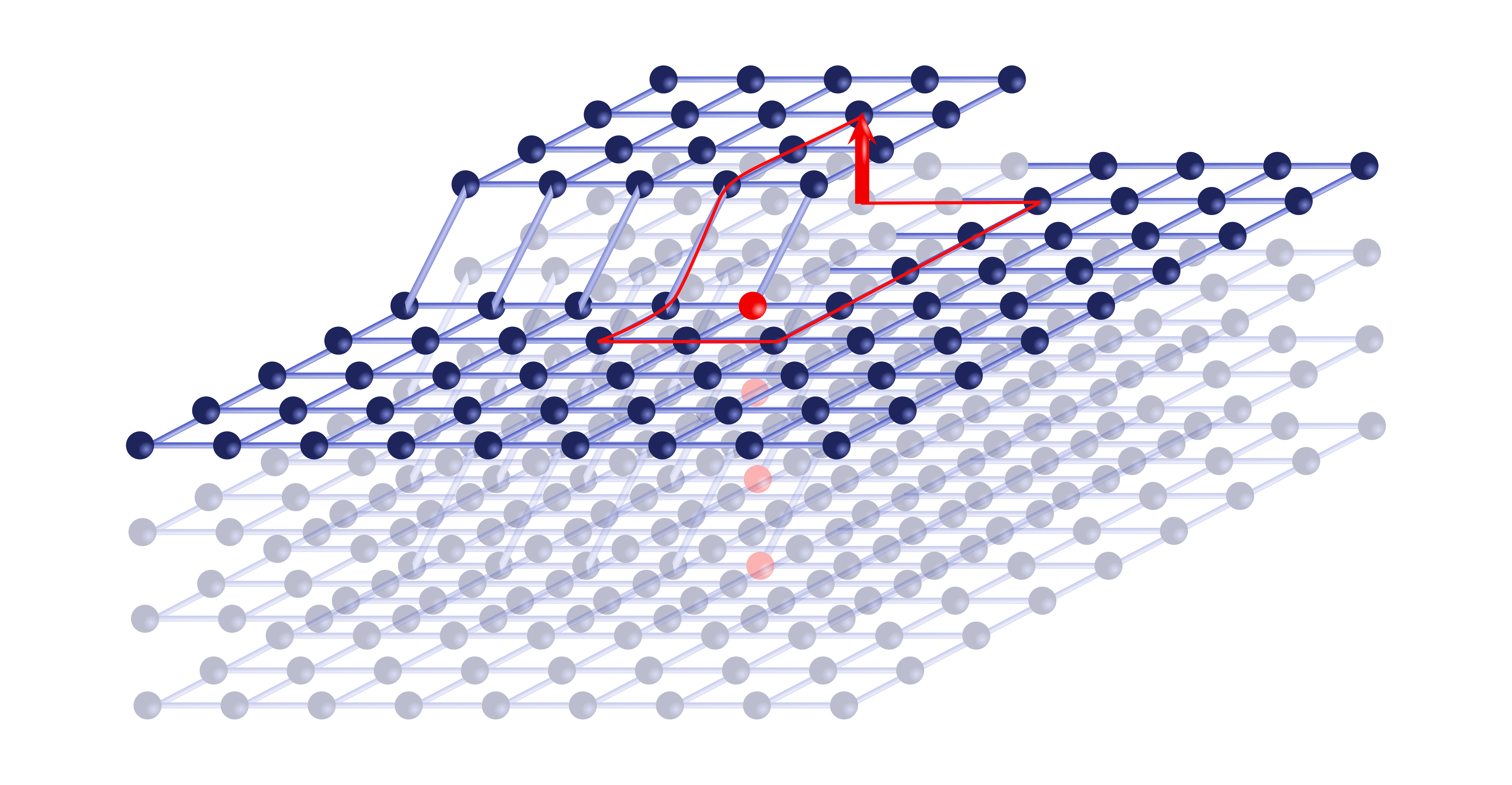}\label{fig:screw dislocation}}
\caption{Dislocation lines (red spheres) in the relativistic 3D ``world crystal" (two space and one time direction), formed by insertion of a half-plane of particles. Shown in red is the contour that measures the mismatch quantized in the Burgers vector (red arrow). If the Burgers vector is orthogonal to the dislocation line it is an edge location; if the Burgers vector is parallel it is a screw dislocation. In non-relativistic 2+1D there are only edge dislocations, since the Burgers vector is always purely spatial.}
\end{figure}

These complications become manifest considering the topological defects: the dislocations and disclinations.  The 
dislocation is the topological defect associated with the restoration of the translations. The edge dislocation can be 
viewed as the insertion of a half-plane of extra atoms terminating at the dislocation core. One immediately
infers that it carries a vectorial topogical charge: the Burgers vector indexed according to the Miller indices of the 
crystal. In 2+1 dimensions the dislocation is a particle (like the vortex) and as an extra complication the Burgers
vector can either lie perpendicular (``edge dislocation", Fig. \ref{fig:edge dislocation}) or parallel (``screw dislocation", Fig. \ref{fig:screw dislocation}) to the propagation 
direction of its world line.  The disclination is on the other hand associated with the restoration of the rotational 
symmetry. This can be obtained by the Volterra construction: cut the solid, insert a wedge and glue together
the sides (see Fig. \ref{fig:disclination}). This carries also a vectorial charge (the Frank vector). Finally dislocations and disclinations are not
independent. On the one hand, the disclination can be viewed as a stack of dislocations with parallel Burgers 
vectors, while the dislocation can be viewed as a disclination--antidisclination pair displaced by a lattice constant.

\begin{figure}
 \includegraphics[height=2.6cm]{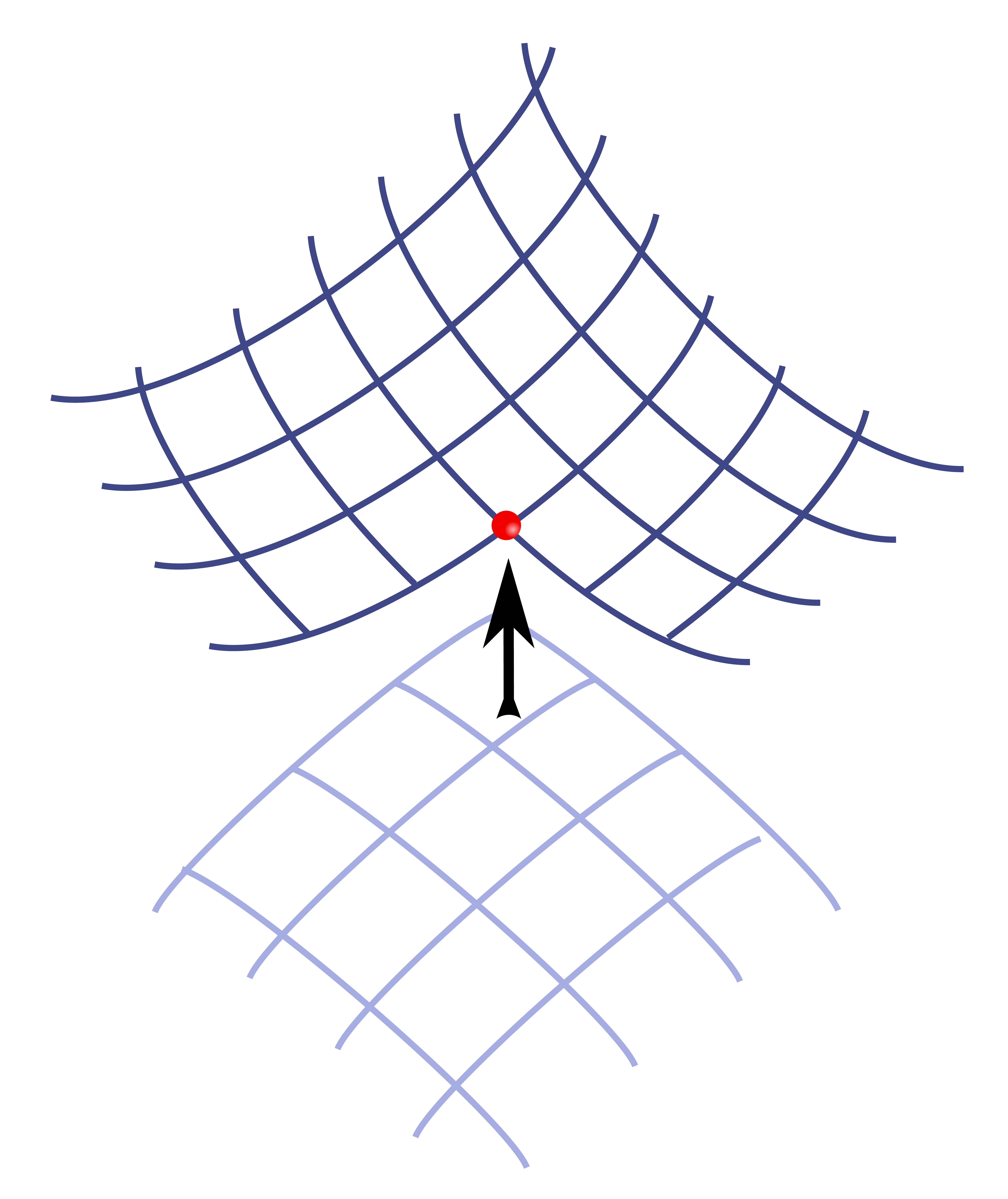}
\caption{90$^\circ$ disclination in a square lattice. A wedge is inserted into a cut in the lattice. There is now one lattice point with five instead of four neighboring sites (red); going in contour around this point will result in an additional 90$^\circ$ rotation. The associated topological charge is the Frank vector, orthogonal to the plane and of size 90$^\circ$. As the dislocation, in 2+1D spacetime the disclination point will trace out a world line.\label{fig:disclination}}.
\end{figure}

Dislocations and disclinations have however a different identity and this makes possible to give a tight, 
topological definitions of quantum  smectic and nematic order. A state where dislocations have spontaneously 
proliferated and condensed, while the disclinations are still massive, is a quantum liquid crystal. Since a disclination
is coincident with a ``uniform magnetization" of Burgers vectors, one cannot have a net density of parallel Burgers vectors as long as disclinations are suppressed. The Burgers vectors of the dislocations in the 
condensate have to be anti-parallel and therefore the dislocation breaks orientations rather than rotations, with 
the ramification that the order parameter is a director instead of a vector. Finally, when all orientations of the Burgers vectors are 
populated equally in the condensate one deals with a nematic breaking only space rotations. When only a particular 
Burgers vector orientation is populated one is dealing with a smectic because the translations are only restored in
the direction of the Burgers vector: the system is in one direction a superfluid and in  the other still a solid. 
To complete this outline, when the coupling constant is further increased there is yet another quantum phase 
transition associated with the proliferation of disclinations turning the system into an isotropic superfluid. 

Let us now review the ``dislocation duality": in close analogy with vortex duality, this shows how crystals and 
liquid crystals are related via a weak--strong duality. The requirement that disclinations have to be kept out
of the vacuum is actually a greatly simplifying factor. One  follows the same dualization procedure for the dislocations
as for the vortices. Hence, we introduce Hubbard--Stratonovich auxiliary tensor fields $\sigma_{\mu \nu}$,
rewriting the action as,

\begin{equation}
S = \int d\tau dx^2\ \left[ \frac{1}{4\mu} \left( \sigma_{\mu \nu}^2 - \frac{\nu}{1 + \nu} \sigma_{\mu \mu}^2 \right)
+ i \sigma_{\mu \nu} w_{\mu \nu} \right]\;,
\label{stressact}
\end{equation}

where $\nu = \lambda /2(\lambda + \mu)$ is the Poisson ratio. We  divide the displacement fields
(having the same status as the phase field in vortex duality) in 
smooth and multivalued parts $u_{\mu} = u_{\mu}^{\mathrm{smooth}} + u_{\mu}^{\mathrm{MV}}$, and integrating out the smooth 
strains yields a Bianchi identity,

\begin{equation}
\partial_{\mu} \sigma_{\mu \nu} = 0\;,
\label{stresscons}
\end{equation}

The physical meaning of $\sigma_{\mu \nu} $ is that they are the stress fields, which are conserved in the absence
of external stresses as in Eq. (\ref{stresscons}): the above is just the stress--strain duality of elasticity theory. One now
wants to parametrize the stress fields in terms of a  gauge field. Since the stress tensor is symmetric this is 
most naturally accomplished in  terms of Kleinert's double curl gauge fields,

\begin{equation}
\sigma_{\mu\nu} = \epsilon_{\mu \kappa \lambda} \epsilon_{\nu \kappa' \lambda'} \partial_{\kappa} \partial_{\kappa'}
B_{\lambda \lambda'}
\label{double curl}
\end{equation}

while the $B$'s are {\em symmetric} tensors, otherwise transforming as 1-form $U(1)$ gauge fields.
 
To maintain the analogy with the vortex duality as tightly as possible, one can as well parametrize it in a normal gauge 
field, $\sigma_{\mu \nu} = \varepsilon_{\mu \kappa \lambda} \partial_{\kappa} A_{\lambda}^{\nu}$ with the 
requirement that one has to impose the symmetry of the stress tensor explicitly by Lagrange multipliers. Using this
route one finds that the multivalued strains turn into a source term $i A_{\mu}^{\nu} J_{\mu}^{\nu}$ where,

\begin{equation}
J_{\mu\nu} =  \epsilon_{\mu \kappa \lambda} \partial_{\kappa} \partial_{\lambda} u^\mathrm{MV}_{\nu}\;,
\label{dislocationcur}
\end{equation}

This is just like a vortex current carrying an extra ``flavor" $\nu$. This is the dislocation current, where the flavor
indicates the $d$+1 components of the Burgers vector. As the vortices, dislocations have long-range interactions 
which are parametrized by the gauge fields $A$ (or $B$), with the special effect that these are only active in the
directions of the Burgers vectors. 

The double curl gauge fields have the advantage that the symmetry is automatically built in while the ``extra
derivatives" make possible to identify the disclination currents.  One finds,

\begin{equation}
S = \int d\tau dx^2\ \left[ \frac{1}{4\mu} \left( \sigma_{\mu \nu}^2 - \frac{\nu}{1 + \nu} \sigma_{\mu \mu}^2 \right)
+ i B_{\mu \nu} \eta_{\mu \nu} \right]\;,
\label{dualelaction}
\end{equation}

where the ``stress gauge fields" $B$ are sourced by a total ``defect current",

\begin{eqnarray}
\eta_{\mu \nu}  & = & \epsilon_{\mu \kappa \lambda} \epsilon_{\nu \kappa' \lambda'} \partial_{\kappa}
\partial_{\kappa'} w_{\lambda \lambda'}^\mathrm{MV}\;, \nonumber \\
             & =  & \theta_{\mu \nu} - \epsilon_{\mu \kappa \lambda} \partial_{\kappa} J_{\nu \lambda}\;,
\label{defectcur}
\end{eqnarray}

where $\theta_{\mu \nu}$ is the disclination current, and $\nu$ refers to the Franck vector component.  The fact
that the disclination current has ``one derivative less" than the dislocation current actually implies that disclinations
are in the solid confined---in the solid, a disclination is like a quark.

One now associates a much larger core energy to the disclinations than to the dislocations, and upon increasing the
coupling constant a loop blowout transition will occur involving only the dislocation world lines---it is obvious from the
single curl gauge field formulation that dislocations are just like vortices carrying an extra ``Burgers flavor".  
To obtain the quantum nematic one populates all Burgers directions equally and after some straightforward 
algebra one obtains the effective action for the ``Higgsed stress photons" having the same status as 
Eq. (\ref{vmassphot}) for the Mott insulator,  
\begin{widetext}
\begin{equation}
S = \int d\tau dx^2 \ \left[ m^2_\mathrm{nem} \sigma_{\mu \nu} \frac{1}{\partial^2} \sigma_{\mu \nu} +
\frac{1}{4\mu} \left( \sigma_{\mu \nu}^2 - \frac{\nu}{1 + \nu} \sigma_{\mu \mu}^2 \right)
+ i B_{\mu \nu} \theta_{\mu \nu} \right]\;,
\label{nematicaction}
\end{equation}
\end{widetext}
where $\sigma$ should be expressed in the double curl gauge field $B_{\mu \nu}$ according to 
Eq. (\ref{double curl}).  In terms of the regular gauge fields $A_{\mu}^{\nu}$, the first term represents a Higgs mass, 
while the second term is like a Maxwell term. Nevertheless, in the nematic the disclinations still act as sources
coupling to the double curl gauge fields. 

Ignoring the disclinations, one finds in 2+1 D that Eq. (\ref{nematicaction}) describes a state is quite similar to 
a Mott insulator: all excitations are massive, and one finds now a triplet of massive ``photons''. These are counted
as follows: there are two propagating (longitudinal and transversal acoustic) phonons of the background world
crystal, turning into ``stress photons" after dualization and acquiring a mass in the nematic. In  addition, the
dislocation condensate adds one longitudinal stress photon.

As it turns out, the rules change drastically upon breaking the Lorentz invariance. In a crystal formed from 
material bosons, displacements in the time direction $u_{\tau}$ are absent, and this has among others the 
consequence that the dislocation condensate does not couple to compressional stress. Instead of the 
incompressible nature of the relativistic state, one finds now two massles modes in  the quantum nematic:
a rotational Goldstone boson associated with the restoration of the broken rotational symmetry, and a
massless sound mode which can be shown to be just the zero sound mode of the superfluid. The non-relativistic
quantum liquid crystals are automatically superfluids as well and their relation to gravity is obscured.

Turning to the 3+1D case one finds as extra complication that dislocations turn into strings and one has to address
the fact that the ``stress superconductor" is now associated with a condensate of strings.  One meets the same complication
in the case of the vortex duality and we showed recently how to handle this\cite{BeekmanSadriZaanen11}. The outcome is
actually quite straightforward:  the effective London actions of the type Eqs. (\ref{vmassphot},\ref{nematicaction}) have the 
same form regardless whether one deals with particle or string condensates, and these enter through the Higgs term
$\sim \sigma^2 /\partial^2$.  

How to interpret the 2+1D relativistic quantum nematic? There are no low energy excitations and it only reacts to
disclinations. It has actually precisely the same status as a flat Einsteinian spacetime in 2+1D that only feels
the infinitesimal vibrations associated with gravitational events far away. Similarly, using the general relativity (GR) technology of the
next section, it is also straightforward to demonstrate\cite{ZhuJiang11,BeekmanSadriZaanen12} that in 3+1D one ends up with two massless spin-2 modes:
the gravitons.  To prove that it is precisely linearized gravity, let us consider next the rules of Kleinert that allow to explicitly relate these
matters to gravitational physics. 

\subsection{The field theory of quantum elasticity and geometry: the Kleinert rules.}\label{sec:The field theory of quantum elasticity and geometry: the Kleinert rules.}

Elaborating on a old tradition in ``mathematical metallurgy", Kleinert identified an intriguing portfolio of 
general correspondences between the field theory describing elastic media and the geometrical notions 
underlying general relativity. In order to appreciate what comes, we need to familiarize the reader with 
some of the entries of this dictionary. For an exhaustive exposition we refer the reader to Kleinert's books
on the subject \cite{Kleinert89b,Kleinert08}.

GR is a geometrical theory which departs from a metric $g_{\mu \nu}$, such that an infinitisimal 
distance is measured through,
 
\begin{equation}
ds = g_{\mu \nu} dx_{\mu} dx_{\nu}\;,
\label{metric}
\end{equation}

One now insists that the physics  is invariant under local coordinate transformations (general covariance) $x_\mu \to \xi_\mu (x_\nu)$;  infintesimal transformations then are like  gauge transformations of the metric,

\begin{equation}
g_{\mu \nu} \rightarrow g_{\mu \nu} + \partial_\mu \xi_\nu + \partial_\nu \xi_\mu \equiv   g_{\mu \nu}+ h_{\mu \nu}\;,
\label{diffs}
\end{equation}

Only quantities are allowed in the theory which are invariant under these transformations and insisting on
the minimal number of gradients, one is led to the Einstein--Hilbert action governing spacetime,

\begin{equation}
S = - \frac{1}{2 \kappa} \int d^d x dt\ R \sqrt{-g}\;,
\label{hilbert}
\end{equation}

where $g = \det g_{\mu \nu}$ and $R$ the Ricci scalar, while $\kappa$ is set by Newton's constant. Together
with the part describing the matter fields, the Einstein equations follow from the saddle points of this action.

How to relate this to solids? Imagine that one lives inside a solid and all one can do to measure distances is 
to keep track how one jumps from unit cell to unit cell. In this way one can define a metric ``internal" to the 
solid, and the interesting question becomes: what is the fate of the diffeomorphisms  (``diffs'')  Eq. (\ref{diffs})?
In order to change the metric one has to displace the atoms and this means that one has to {\em strain}
the crystal,

\begin{equation}
g_{\mu \nu} \rightarrow g_{\mu \nu} + w_{\mu \nu}\;,
\label{infinintesimal diffs}
\end{equation}

But the strain fields are surely not gauge fields: the elastic energy  Eq. (\ref{elfieldth}) explicitly depends
on the strain. Obviously, the crystal is non-diffeomorphic and it is characterized by a ``preferred" or
``fixed" frame. This is the deep reason that normal crystals have nothing to do with GR.

In standard GR the objects  that are invariant keep track of  curvature and these 
appear in the form of curvature tensors in the Einstein equations. Linearizing these, assuming only
infinitisimal diffs as in Eq. (\ref{diffs}),
one finds for the Einstein tensor appearing in the Einstein equations, say in the 2+1D case to avoid superfluous labels,

\begin{equation}
G_{\mu \nu}   =  \epsilon_{\mu \kappa \lambda} \epsilon_{\nu \kappa' \lambda'} \partial_{\kappa}
\partial_{\kappa'} h_{\lambda \lambda'}
\label{einsteinten}
\end{equation}

One compares this with the disclination current Eq. (\ref{defectcur}) and one discovers that these are the 
same expressions after associating the strains $w_{\mu \nu}$ with the infinitesimal diffs $h_{\mu \nu}$. 
 This is actually no wonder: at stake is that the property of curvature is independent of the gauge choice
for the metric. One can visualize the curved manifold in a particular gauge fix, and this is equivalent to the 
fixed frame. The issue is that curvature continues to exist when one lets loose the metric in the gauge volume. 

What is the meaning of the dislocation tensor? Cartan pointed out to Einstein that his theory was geometrically
incomplete: one has to allow  also for the property of torsion. It turns out that torsion is  ``Cartan-Einstein"
GR sourced by spin currents and the effects of it turn out to be too weak to be observed (see e.g. Ref. \onlinecite{GronwaldHehl96}). In  the present context, the 
torsion tensor appearing in the equations of motions  precisely corresponds with the dislocation currents. 
With regard to these topological aspects, crystals and spacetime are remarkably similar.   

However, given the lack of general covariance the dynamical properties of spacetime and crystals are entirely
different.  For obvious reasons, spacetime does not know about phonons while crystals do not know about
gravitons, let alone about black holes. A way to understand why things go so wrong is to realize that 
the disclinations encode for curvature, while gravitons can be viewed as infinitisemal curvature fluctuations. 
As we already explained, disclinations are confined in crystals meaning that it costs infinite energy to create
curvature fluctuations in normal solids. 

Let us now turn to the relativistic quantum nematics: here the situation looks much better.  Gravity in 2+1D is 
incompressible in the sense that the constraints do not permit massless propagating modes, the gravitons. We also found out that
disclinations are now deconfined and they appear as sources in the effective action Eq. (\ref{nematicaction}):
this substance knows about curvature. In fact, one can apply similar considerations to the 3+1D case, 
where two massless spin-2 modes are present. The relativistic quantum nematic in 3+1D 
behaves quite like spacetime!

To make the identification even more precise, one notices that the expression for the linearized Einstein 
tensor Eq. (\ref{einsteinten}) is coincident with the expression for the stress tensor in terms of the double 
curl gauge field $B_{\mu \nu}$, Eq. (\ref{double curl}).  But now one is dealing with gauge invariance both 
of $B_{\mu \nu}$ and $h_{\mu \nu}$ while they are both symmetric tensors.  At least on the linearized level
the stress tensor \emph{is} the Einstein tensor. It is now easy to show that the Higgs term in the theory of the
nematic when expressed in terms of the linearized Einstein tensors,

\begin{eqnarray}
\sigma \frac{1}{\partial^2} \sigma & = & G   \frac{1}{\partial^2}  G \nonumber \\
& \rightarrow & R
\label{einsteintoricci}
\end{eqnarray} 

actually reduces to the Ricci tensor $R$, demonstrating that one recovers the Einstein-Hilbert action at
distances large compared  to the Higgs scale. Again, this only holds in the linearized theory.  
This works in the same way in 3+1 (and higher) dimensions which is the easy way to 
demonstrate that gravitons have to be present\cite{ZhuJiang11,BeekmanSadriZaanen12}.  At least the 
linearized version of the Einstein--Hilbert action appears to be precisely coincident with the effective 
field theory describing the collective behavior of the quantum nematic!

Although this all looks convincing there is still a gap in the conceptual understanding of what has happened 
with the geometry of the crystal in the presence of condensed dislocations. The emergence of gravity 
requires that the original spacetime defined by the crystal has to become diffeomorphic. The fields as
of relevance to the dynamics of the nematic are healthy in this regard but they belong to the dual side.
The analogy with the Mott insulator is now helpful: to demonstrate that gravity has emerged requires the 
demonstration that the spacetime of the original crystal is diffeomorphic and that is equivalent to
demonstrating that in the vortex condensate the superfluid phase acquires a compact $U(1)$ gauge invariance.
The diffeomorphic nature of the stress gauge fields telling about the excitations of the quantum nematic 
has in turn the same status as the gauge fields that render the vortex condensate to be a superconductor. 

\begin{figure*}

\huge
 $\frac{1}{\sqrt{2}}\Big( |$  \raisebox{-11mm}{\includegraphics[height=25mm]{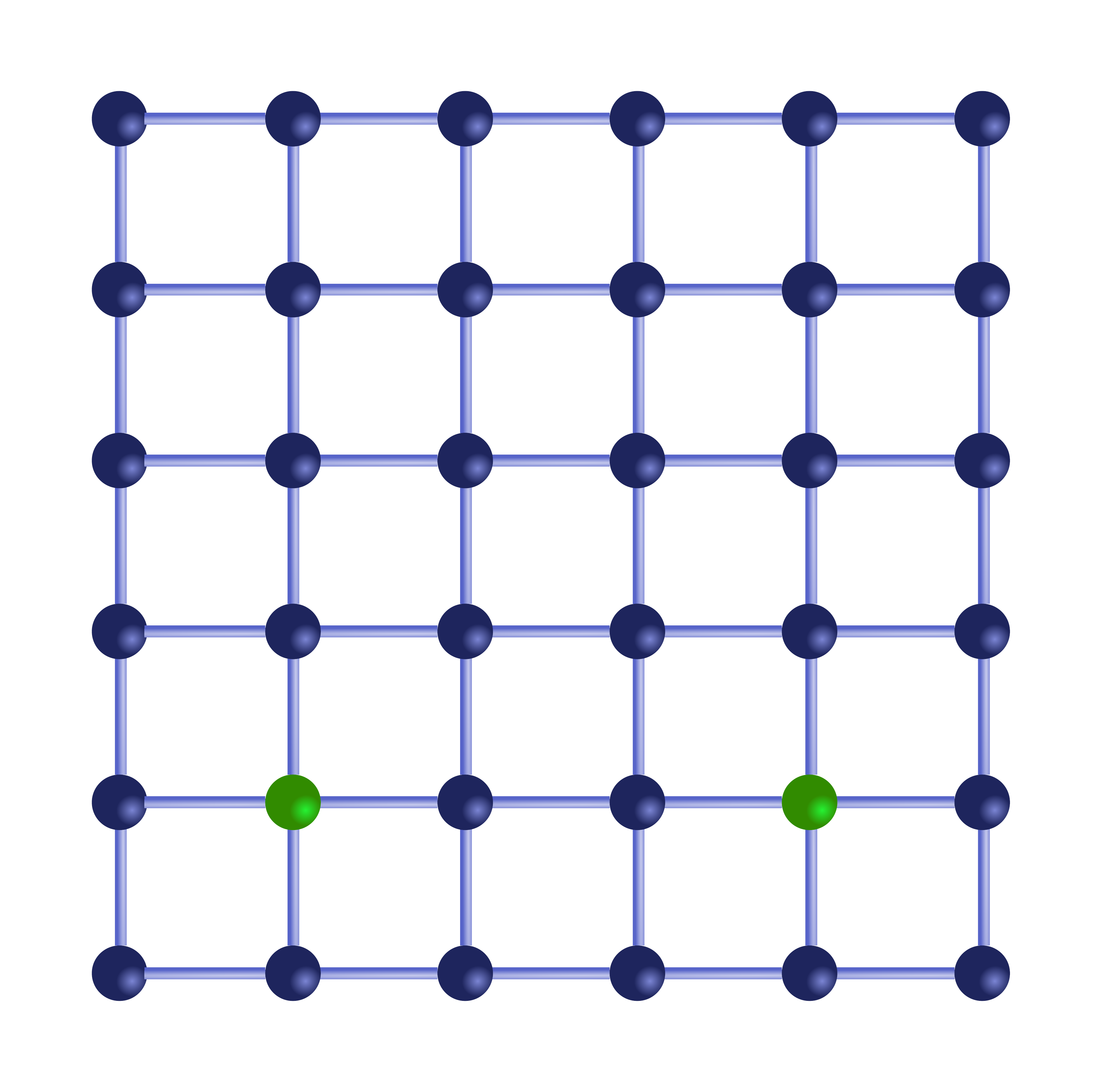}} $\rangle + |$  \raisebox{-11mm}{\includegraphics[height=25mm]{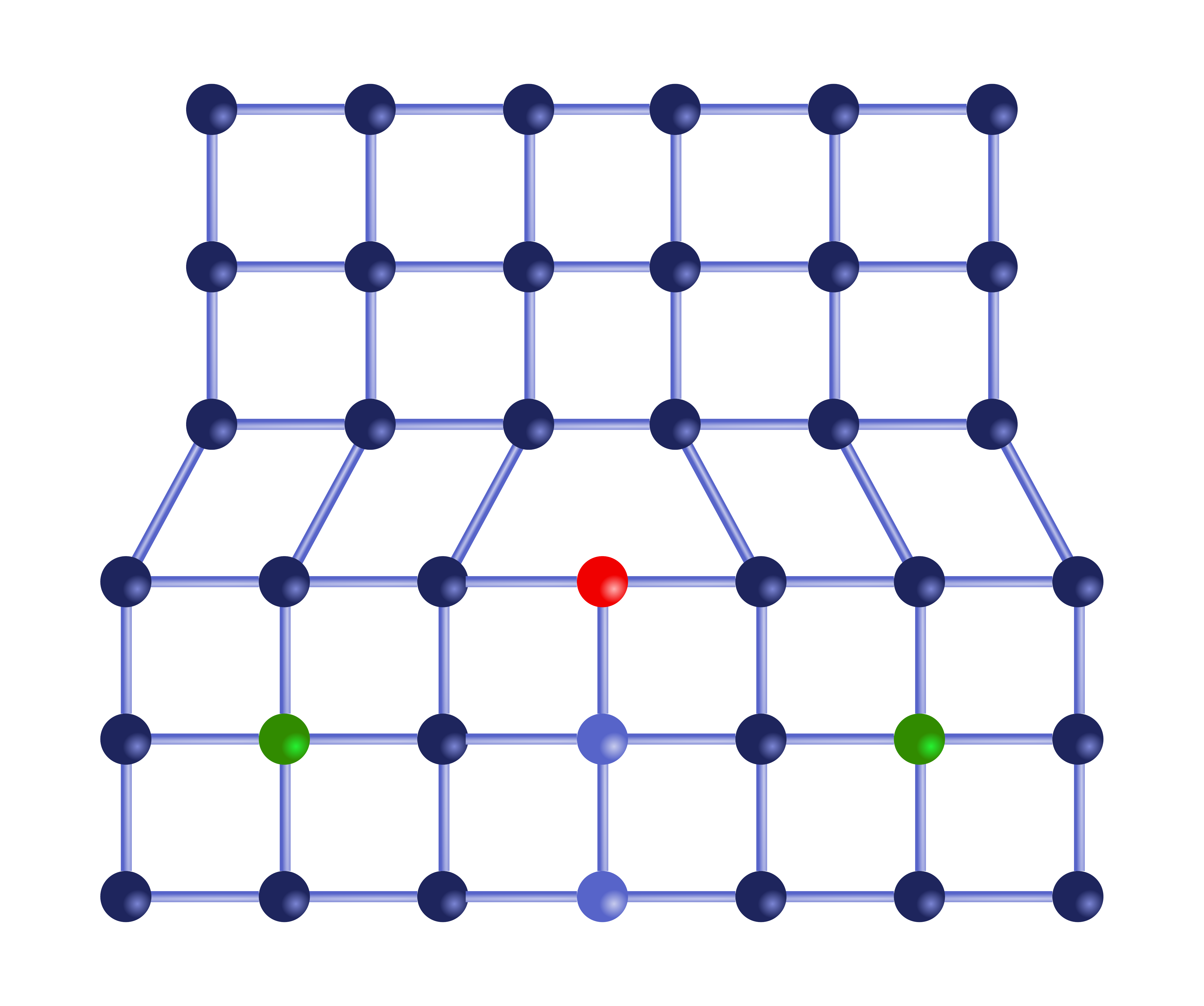}} $\rangle \Big)$
\normalsize

\caption{In the dislocation condensate (quantum nematic), the distance between two points (green dots) is in a coherent 
superposition of having zero, one or any number of half-line insertions (light blue) or dislocations (red dot) in between them, and therefore the number of lattice spacings in between them is undefined. This is equivalent to having the Einstein translations fully gauged: there is a diffeomorphism between configurations with any number of lattice spacings in between the two points.}\label{fig:dislocation superposition}
\end{figure*}

The good news is that we can use the same ``first quantization" trick that helped us to understand the 
emergence of the stay-at-home gauge in the vortex condensate to close this conceptual gap. As for the vortices,
it is easy to picture what happens to the metric of the crystal when the coherent superpositions of dislocation
configurations associated with the dual stress superconductor are present. Let us repeat the exercise at the
end of the previous section, by comparing how two points some distance apart communicate with each other,
but now focusing on the metric properties. This is illustrated in Fig. \ref{fig:dislocation superposition}: imagine that no dislocation is 
present between the two points and one needs $N$ jumps to get from one point to the other. However, this 
configuration is at energies less than the Higgs mass of the quantum nematic necessarily in coherent 
superposition with a configuration where a dislocation has moved through the line connecting the
two points: one now needs $N+1$ hops and  since these configurations are in coherent superposition
``$N = N+1$" and the geometry is now truly diffeomorphic!

However, there is one last caveat. Although translational symmetry is restored in the quantum nematic, the rotations
are still in a fixed frame and even spontaneously broken! This is different from full Einstein gravity: in real spacetime also the Lorentz transformations (rotations in our Euclidean setting) are fully gauged. In order to understand this point, let us start from special relativity, which has the global symmetry of the Poincar\'e group comprising translations and Lorentz transformations. The translations form a subgroup, such that translational and rotational symmetry are easily distinguishable. More precisely, the generators of translations are ordinary derivatives $\partial_\mu$ which commute $[\partial_\mu , \partial_\nu]=0$.   In many ways, going from special to general relativity is from going from global to local Poincar\'e symmetry\cite{GronwaldHehl96}. Indeed, referring to elasticity language, it seems to make sense to restore first translational and then rotational symmetry, ending up in a perfectly locally symmetric ``liquid'' state. 

However, it has long been known that such ``gauging of spacetime symmetry'' is very intricate, which has to do with the definition of locality under such transformations. What happens is that local coordinate tranformations of the form $x_\mu \to \xi_\mu (x_\nu)$, which are in fact local translations, can also correspond to local rotations. The local translations no longer form a subgroup, as the generators of translations should be augmented to those of \emph{pa\-ral\-lel translations}, defined by \cite{HehlVonderheydeKerlickNester76},
\begin{equation}
D_\mu = \partial_\mu +  \Gamma_\mu^{\phantom{\mu}\kappa\lambda} f_{\kappa\lambda},
\end{equation}
where $\Gamma_\mu^{\phantom{\mu}\kappa\lambda}$ is the connection and  $ f_{\kappa\lambda}$ is the generator of local rotations. 
Such modified derivatives do not commute, and two consecutive translations may result in a finite rotation. Such symmetry structure is actually at the heart at everything non-linear
happening in  Einstein theory including black holes.

Going back to what we now know of the quantum nematic, it is clear that it cannot correspond to full GR, since rotational symmetry as reflected by disclinations is still gapped. Nevertheless, the identification between quantum nematics and {\em linearized} gravity is in perfect
shape. Linearized gravity is a special and somewhat pathological limit of full GR, as it only applies to nearly globally Lorentz symmetric systems. It was quite some time ago
realized  that such systems are symmetric under global Lorentz transformations and infinitesimal coordinate transformations (see ch. 18,35 in Ref. \onlinecite{MisnerThorneWheeler73}). This is equivalent to fixing the 
Lorentz frame globally yet allow for infinitesimal Einstein translations. Under such conditions the equations of motion of linearized gravity follow automatically. 
 
Here we have demonstrated that linearized gravity---a very peculiar limiting case of GR---is actually literally 
realized in a quantum nematic. The deeper reason is that in a quantum nematic the rotational symmetry of 
(Euclidean) spacetime is global and even spontaneously broken, while the restoration of the translational 
symmetry by the dislocation condensate has caused the fixed frame internal coordinate system of the crystal 
to turn into a geometry that is characterized by a covariance exclusively associated with infinitesimal 
translational coordinate transformations.

\section{Conclusions.}

In so far as the Abelian-Higgs (or ``vortex") duality is concerned we have presented here no more than 
a clarification. Living on the ``dual side", where the Bose-Mott insulator appears as just a relativistic
superconductor formed from vortices,  the  emerging stay-at-home local charge conservation 
from the canonical representation in terms of the Mott insulating phase of the  
Bose-Hubbard model is not manifestly recognizable. However, the dual vortex language contains all the 
information required to reconstruct precisely the nature of the field configurations  of the ``original" superfluid
phase fields which are realized in the vortex superconductor. By inspecting these we identified a very
simple but intriguing principle.   The local charge conservation of the Mott insulator, associated 
with the emergent stay-at-home compact $U(1)$ gauge symmetry, is  generated in the 
vortex condensate by the quantum mechanical principle that states in coherent superposition
``are equally true at the same time"---the Schr\"odinger cat motive.  

We find this simple insight useful since it yields a somewhat more general view on the nature
of strong--weak dualities. We already emphasized that Mott insulators as defined through the local 
conservation of charge do not necessarily need a lattice. One does not have to dig deep to find an example:
our dual superconductor is just a relativistic superconductor in 2+1D, which is in turn dual to a
Coulomb phase that can also be seen as a superfluid. The charge associated with this superfluid is
locally conserved in the superconductor, regardless of whether the superconductor lives on a lattice or
in the continuum.    

We find the emergent gauging of translational symmetry realized in the quantum 
nematic an even better example of the usefulness of this insight. Earlier work indicated
that the relativistic version of this nematic is somehow associated with emergent gravity. Resting on
the ``coherent superposition" argument it becomes directly transparent what causes the gauging of the 
crystal coordinates: the condensed dislocations ``shake the coordinates coherently" such that infinitesimal 
Einstein translations appear while the Lorentz frame stays fixed. This emergent symmetry imposes that the 
collective excitations of the quantum nematic have to be in one-to-one correspondence with linearized   
gravity. 

Our message is that we have identified a mechanism for the ``dynamical generation" of gauge symmetry
which is very simple but also intriguing viewed from a general physics perspective: the quantum mechanics principle
of states in coherent superposition being ``equally true at the same time" translates directly to the principle that the
global symmetry that is broken in the ordered state is turned into a gauge symmetry on the disordered side just by the 
quantum un-determinedness of the topological excitations in the dual condensate. This raises the interesting question:
is quantum coherence required for the emergence of local symmetry, or  can it also occur in classical systems?

This question relates directly to the spectacular discovery of ``Dirac monopoles" in spin ice\cite{BramwellGingras01}. Castelnovo, Moessner and
Sondhi\cite{CastelnovoMoessnerSondhi07} realized that the manifold of ground states (``frustration volume") of this classical geometrically frustrated spin problem is coincident 
with the gauge volume of a compact $U(1)$ gauge theory, with the ramification that it carries Dirac monopoles as 
topological excitations. All along it has been subject of debate to what extent these monopoles can be viewed as 
literal Dirac monopoles in the special ``vacuum" realized in the spin ice, or rather half-bred cartoon versions of the real thing.
With our recipe at hand it is obvious how to make them completely real: imagine the classical spin ice to fill up Euclidean 
spacetime, and after Wick rotation our ``coherent superposition principle" would have turned the  frustration volume of the classical
problem into a genuine gauge volume since by quantum superposition all degenerate states would be ``equally true at the same time".

The ambiguity associated with the classical spin ice monopoles is rooted in the role of time. In principle, by doing time-resolved measurements
one can observe every particular state in the frustration volume and this renders these states to be not gauge equivalent. However,
all experiments which have revealed the monopoles involved large, macroscopic time scales. One can pose the question whether it is 
actually possible under these conditions to define observables that can discriminate between the ``fake" monopoles of spin ice and the monopoles of Dirac. Perhaps
the answer is pragmatic: as long as ergodicity is in charge, one can rely on the ensemble average instead of the time average, and as long as
the time scale of the experiment is long enough such that one is in the ergodic regime, the frustration volume will ``disappear" in the ensemble average.
For all practical purposes  one is then dealing with a genuinely emergent gauge symmetry which tells us that in every regard the spin ice monopole 
is indistinguishable from the Dirac monopole.

\acknowledgments  
{\em Acknowledgements.}
We thank H. Kleinert, D. Sadri, V. Cvetkovic, Y.-W. Sun, T. Senthil  and Y. Liu for helpful discussions. This work is supported by the 
Nederlandse Organisatie voor Wetenschappelijk Onderzoek (NWO) via a Spinoza grant.

\bibliography{ref_emgauge}

\end{document}